\documentclass[conference]{IEEEtran}
\IEEEoverridecommandlockouts

\usepackage{color}
\usepackage{graphicx}
\usepackage{amsmath}
\usepackage{amssymb}

\newcommand{\nid}{\noindent}
\newcommand{\non}{\nonumber}

\renewcommand{\thesubsubsection}{\thesubsection.\arabic{subsubsection}}

\title{Hybrid Procoder and Combiner Design for Secure Transmission in mmWave MIMO Systems}

\author{ \IEEEauthorblockN{Xiaowen Tian$^{\dag}$, Ming Li$^{\dag}$, Zihuan Wang$^{\dag}$, and Qian Liu$^{\ddag}$
\vspace{-0.0 cm} }\\
\IEEEauthorblockA{$^{\dag}$School of Information and Communication Engineering   \\  Dalian University of Technology, Dalian, Liaoning 116024, China  \\ E-mail: \texttt{\{tianxw,wangzihuan\}@mail.dlut.edu.cn, mli@dlut.edu.cn}}

\IEEEauthorblockA{$^{\ddag}$School of Computer Science and Technology  \\  Dalian University of Technology, Dalian, Liaoning 116024, China\\ E-mail: \texttt{qianliu@dlut.edu.cn}} }

\begin{document}
\pagestyle{empty}

 \maketitle

\begin{abstract}
Millimeter wave (mmWave) communications have been considered as a key technology for future 5G wireless networks. In order to overcome the severe propagation loss of mmWave channel,  mmWave multiple-input multiple-output (MIMO) systems with analog/digital hybrid precoding and combining transceiver architecture have been widely considered.
However, physical layer security (PLS) in mmWave MIMO systems and the secure hybrid beamformer design have not been well investigated.
In this paper, we consider the problem of  hybrid precoder and combiner design for secure transmission in mmWave MIMO systems in order to protect the legitimate transmission from eavesdropping.
When eavesdropper's channel state information (CSI) is known, we first propose a joint analog precoder and combiner design algorithm which can prevent the information leakage to the eavesdropper.
Then, the digital precoder and combiner are computed based on the obtained effective baseband channel to further maximize the secrecy rate.
Next, if prior knowledge of the eavesdropper's CSI is unavailable, we develop an artificial noise (AN)-based hybrid beamforming approach, which can jam eavesdropper's reception while maintaining the quality-of-service (QoS) of intended receiver at the pre-specified level.
Simulation results demonstrate that our proposed algorithms offer significant secrecy performance improvement compared with other hybrid beamforming algorithms.
\end{abstract}


\vspace{-0.0 cm}
\section{Introduction}

\pagestyle{empty}

Millimeter wave (mmWave) communications, which can provide orders-of-magnitude wider bandwidth than current cellular bands, has been considered as a key technology for future 5G wireless networks \cite{Rappaport IA 13}. The smaller wavelength of mmWave signals enables a large antenna array to be packed in a small physical dimension at the transceiver ends.
However, conventional full-digital precoder and combiner are realized using a large number of expensive radio frequency (RF) chains and energy-intensive analog-to-digital converters (ADCs), which are impractical in the mmWave communication systems. Recently, economic and energy-efficient analog/digital hybrid precoding and combining transceiver architecture has emerged as a promising solution in mmWave multiple-input multiple-output (MIMO) systems.

The hybrid beamforming structure applies a large number of analog phase shifters (PSs) to implement high-dimensional analog beamformer and a small number of RF chains for low-dimensional digital beamformer to provide the necessary flexibility to perform multiplexing/multiuser transmission \cite{Wang CL}. The existing hybrid beamforming designs can be categorized into \textit{i}) codebook-based scheme in which the analog beamformer is selected from certain candidate vectors, such as array response vectors of the channel and discrete fourier transform (DFT) beamformers \cite{SSP}-\cite{Gao TVT 16}; \textit{ii}) codebook-free scheme in which the infinite resolution of PSs is assumed \cite{PE AltMin}, \cite{Gao JSAC 16}. Currently, the codebook-based beamforming designs are more popular because of the less complexity and satisfactory performance due to the special structure of hybrid beamformer and the characteristic of mmWave MIMO channels.

While existing hybrid beamforming designs focus on improving spectral efficiency of a point-to-point mmWave MIMO channel, however, the secrecy in a mmWave MIMO wiretap channel and the beamforming design for the secure transmission have not been well investigated. In recent years, physical layer security (PLS) has been identified as a promising strategy for secure wireless communications. Especially, beamforming technology becomes a powerful tool for enhancing the physical layer security in conventional MIMO systems \cite{GED}, \cite{GSVD}. With the spatial degrees of freedom (DoF) provided by multiple antennas, the transmitter can adjust its beamforming orientation  to reduce/prevent the information leakage to eavesdroppers or generate artificial noise (AN) to jam potential eavesdroppers.
However, the obtained results cannot be directly applied to mmWave MIMO systems due to the different propagation characteristics and the special hybrid beamforming architecture. Therefore, secure transmission in mmWave MIMO systems attracts new research interests \cite{Wanghm AN2}-\cite{Wanghm AN1}.

In \cite{Wanghm AN2}, the network-wide PLS performance of a mmWave cellular network was investigated under a stochastic geometry framework.
In \cite{Wanghm BF}, the authors considered a mmWave system with the multi-input single-output (MISO) channel and presented two simple beamformer designs for the secure transmission.
Based on this system model, in \cite{Wanghm AN1} the authors further introduced a new form of AN generation method depending on the propagation characteristics of the mmWave channel. Unfortunately, all those mentioned works focus on the comprehensive secrecy performance analysis rather than the beamformer design.
More importantly, the presented simple beamformer designs are based on the conventional full-digital beamforming architecture, which are not practical in the mmWave MIMO systems comparing with the hybrid beamforming structure.
To the best of our knowledge, the hybrid beamformer design for the secure transmission in the mmWave MIMO systems has not been studied yet.

In this paper, we investigate hybrid beamformer design for the secure transmission and propose a novel codebook-based hybrid precoder and combiner design algorithm in order to protect the legitimate transmission from eavesdropping.
In the case that eavesdropper's channel state information (CSI) is available, we first develop a joint analog precoder and combiner design algorithm which can prevent the information leakage to the eavesdropper.
Then, the digital precoder and combiner are computed based on the obtained effective baseband channel to further maximize the secrecy rate.
Then, when prior knowledge of the eavesdropper's CSI is unavailable, we introduce an AN-based hybrid beamforming approach, which can generate disturbance to the eavesdropper while maintaining the quality-of-service (QoS) of the intended receiver at the pre-specified level.
Simulation results demonstrate the significant secrecy performance improvement of our proposed algorithms compared with other hybrid beamforming algorithms.

%

\section{System Model and Problem Formulation}

\subsection{System Model}
\label{sec: model}

\begin{figure}[!t]
\centering
\includegraphics[width= 3.5 in]{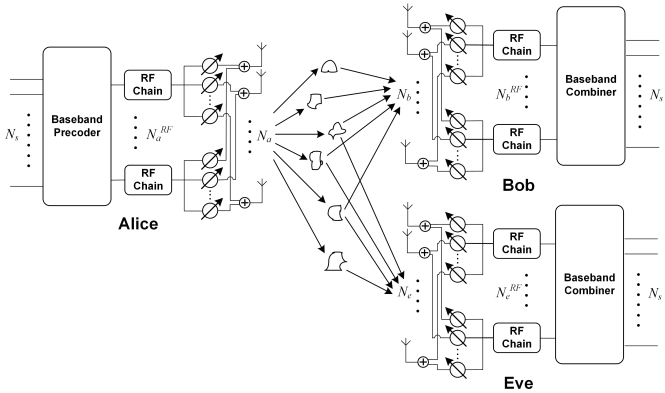}\vspace{-0.3 cm}
\caption{The mmWave MIMO wiretap system with hybrid precoder and combiners. The channel is modeled as scatterer-sharing model.}
\label{fig:system model}\vspace{-0.3 cm}
\end{figure}

We consider a mmWave MIMO wiretap system as illustrated in Fig. \ref{fig:system model}, in which the legitimate transmitter Alice is equipped with $N_a$ antennas and $N^{RF}_a$ RF chains to simultaneously transmit $N_s$ data streams to the legitimate receiver Bob, who is equipped with $N_b$ antennas and $N^{RF}_b$ RF chains.
To ensure the efficiency of the communication with the limited number of RF chains, we assume $N_{RF} = N_a^{RF} = N_b^{RF}$ and  the number of data streams is constrained as $N_s \leq N_{RF}$.
There also exists an eavesdropper Eve, who is equipped with $N_e$ antennas and $N^{RF}_e$ RF chains, $N_s \leq N^{RF}_e$, and attempt to overhear the data transmission from Alice to Bob.

The transmitted symbols are firstly processed by an $N_{RF}\times N_s$ baseband digital precoder $\mathbf{F}_{BB}$, then up-converted to the RF domain via $N_{RF}$ RF chains before being precoded with an $N_a \times N_{RF}$ analog precoder $\mathbf{F}_{RF}$. While the baseband precoder $\mathbf{F}_{BB}$ enables both amplitude and phase modifications, the analog precoder $\mathbf{F}_{RF}$ is implemented by analog components like phase shifters and its elements are constrained to satisfy with constant magnitude. In the codebook-based precoder scheme, the analog beamformer is
selected from a pre-specified codebook $\mathcal{F}$, i.e. a set of $N_a \times 1$ vector with constant-magnitude entries. The normalized power constraint is given by $\|\mathbf{F}_{RF}\mathbf{F}_{BB}\|^2_F = N_s$.
Therefore, the transmit signal has a form of
\begin{equation}
\mathbf{x} = \sqrt{P}\mathbf{F}_{RF}\mathbf{F}_{BB}\mathbf{s}
\end{equation}
where $P$ is the transmitted power and $\mathbf{s}$ is the $N_s \times 1$ symbol vector such that $\mathbb{E}\{\mathbf{s}\mathbf{s}^H\}= \frac{1}{N_s}\mathbf{I}_{N_s}$.

Denote $\mathbf{H}_b\in \mathbb{C}^{N_b\times N_a}$ as the channel matrix of Alice-to-Bob channel. Let $\mathbf{n}_b\in \mathbb{C}^{N_b\times 1}$   represent the noise vector of independent and identically distributed (i.i.d.) $\mathcal{CN}(0,\sigma^2_b)$ elements.
At Bob, the received signal \begin{equation} \mathbf{r}_b = \mathbf{H}_b \mathbf{x} + \mathbf{n}_b \end{equation} are first processed by the analog combiner of Bob $\mathbf{W}_{RF,b}$, then the digital combiner $\mathbf{W}_{BB,b}\in \mathbb{C}^{N_{RF} \times N_s}$. Let $\mathbf{W}_b \triangleq \mathbf{W}_{RF,b}\mathbf{W}_{BB,b}$ and $\mathbf{F}  \triangleq \mathbf{F}_{RF }\mathbf{F}_{BB}$ for simplicity. Thus,  the processed receive signal at Bob can be expressed as
 \begin{equation}
\mathbf{\hat{s}}_b = \sqrt{P}\mathbf{W}^H_b \mathbf{H}_b \mathbf{F} \mathbf{s} + \mathbf{W}^H_b \mathbf{n}_b,
\end{equation}
where the subscript ``$b$'' indicates Bob.

The signal processing at Eve have the same procedure. Let $\mathbf{H}_e\in \mathbb{C}^{N_e\times N_a}$ denote the Alice-to-Eve eavesdropping channel and let $\mathbf{n}_e\in \mathbb{C}^{N_e\times 1}$ represent the noise vectors of i.i.d.  $\mathcal{CN}(0,\sigma^2_e)$ elements. Then, we have the processed receive signal at Eve
\begin{equation}
\mathbf{\hat{s}}_e = \sqrt{P}\mathbf{W}^H_e \mathbf{H}_e \mathbf{F} \mathbf{s} + \mathbf{W}^H_e \mathbf{n}_e,
\end{equation}
where the subscript ``$e$'' indicates Eve.

\subsection{MmWave MIMO Channel Model}
The mmWave MIMO channel can be described with the widely used limited scattering channel, in which the number of scatters is $L$, and each scatter is further assumed to contribute a single propagation path between the transmitter and the receiver.
Under this model, the Alice-to-Bob channel matrix $\mathbf{H}_b$ can be expressed as  \vspace{-0.1 cm}
\begin{equation} 
\mathbf{H}_b = \sqrt{\frac{N_a N_b}{\rho_b}}\sum_{l_b=1}^{L_b} \alpha_{l_b} \mathbf{a}_b(\theta_{l_b})\mathbf{a}^H_a(\phi_{l_b}), \vspace{-0.1 cm}
\end{equation}
where
$\rho_b$ denotes the average path-loss between Alice and Bob, and $\alpha_{l_b}$ is the complex gain of the $l_b$-th path and assumed to be Rayleigh distributed.
The variables $\theta_{l_b}\in [0,2\pi]$ and $\phi_{l_b}\in [0,2\pi]$ are the $l_b$-th path's azimuth angles of departure or arrival (AoDs/AoAs) of the transmitter and the receiver, respectively.
$\mathbf{a}_a(\phi_{l_b})$ and $\mathbf{a}_b(\theta_{l_b})$ are the antenna array response vectors at the transmitter and the receiver, respectively.
In this paper, we assume the transmitter and the receivers adopt uniform linear arrays (ULA) for simplicity and then $\mathbf{a}_a(\phi_{l_b})$ and $\mathbf{a}_b(\phi_{l_b})$ are given by  \vspace{-0.1 cm}
\begin{small}
\begin{eqnarray}
\hspace{-0.2 cm}\mathbf{a}_a(\phi_{l_b}) \hspace{-0.3 cm} &  = & \hspace{-0.3 cm} \frac{1}{N_a} [1,e^{j(2\pi/\lambda)d\sin(\phi_{l_b})},...,e^{j(N_a-1)(2\pi/\lambda)d\sin(\phi_{l_b})}]^T, \label{eq: codebook f}\\
\hspace{-0.2 cm} \mathbf{a}_b(\phi_{l_b}) \hspace{-0.3 cm}  &  = & \hspace{-0.3 cm} \frac{1}{N_b} [1,e^{j(2\pi/\lambda)d\sin(\theta_{l_b})},...,e^{j(N_b-1)(2\pi/\lambda)d\sin(\theta_{l_b})}]^T, \label{eq: codebook w}  \vspace{-0.1 cm}
\end{eqnarray}
\end{small}
\nid where $\lambda$ is the signal wavelength and $d$ is the distance between antenna elements. The Alice-to-Eve channel matrix can be written in a similar fashion as  \vspace{-0.1 cm}
\begin{equation}
\mathbf{H}_e = \sqrt{\frac{N_a N_e}{\rho_e}}\sum_{l_e=1}^{L_e} \alpha_{l_e} \mathbf{a}_e(\theta_{l_e})\mathbf{a}^H_a(\phi_{l_e})  \vspace{-0.1 cm}
\end{equation}
\nid with different AoAs $\theta_{l_e}$ and AoDs $\phi_{l_e}$.

Due to the sparse property of mmWave MIMO channel, mmWave communication is usually considered as more secure than conventional MIMO systems since the generated beamformer is too narrow to be eavesdropped if the Eve is not close to Bob.
However, it has been verified that rough surface and tiny building
cracks can cause diffuse scattering in mmWave channel and the diffuse range increases as the wavelength shrinks \cite{scatter sharing channel}.
Therefore, it is highly possible that different receivers share some common scatterers, as shown in Fig. \ref{fig:system model}.
In other words, when Bob and Eve have similar AoDs with some common scatterers and Alice use those AoDs to transmit the secret information to Bob, Eve will have chance to receive very strong signal from Alice, resulting in severe information leakage.
Therefore, it is easier for Eve to eavesdrop secret information in this scatterer-sharing model and we aim to investigate the physical layer security for the mmWave MIMO systems with the scatterer-sharing model.

\subsection{Problem Formulation}

In the context of physical layer security, the secrecy rate is usually used as the performance metric: \vspace{-0.2 cm}
\begin{equation}
R_s = [\log_2\det(\mathbf{I}_{N_s}+\mathbf{S}_b)-\log_2\det(\mathbf{I}_{N_s}+\mathbf{S}_e)]^+ \vspace{-0.2 cm}
\end{equation}
where  \vspace{-0.2 cm}
\begin{eqnarray}
\mathbf{S}_b &=& \frac{P}{N_s} \mathbf{R}^{-1}_{n,b} (\mathbf{W}_{BB,b})^H (\mathbf{W}_{RF,b})^H \mathbf{H}_b \mathbf{F}_{RF} \mathbf{F}_{BB} \non \\
&  & \times \mathbf{F}^H_{BB} \mathbf{F}^H_{RF} \mathbf{H}^H_b \mathbf{W}_{RF,b} \mathbf{W}_{BB,b},\\
\mathbf{S}_e &=& \frac{P}{N_s} \mathbf{R}^{-1}_{n,e} (\mathbf{W}_{BB,e})^H (\mathbf{W}_{RF,e})^H \mathbf{H}_e \mathbf{F}_{RF} \mathbf{F}_{BB} \non \\
&  & \times \mathbf{F}^H_{BB} \mathbf{F}^H_{RF} \mathbf{H}^H_e \mathbf{W}_{RF,e} \mathbf{W}_{BB,e},\\
\mathbf{R}_{n,b} &= &\sigma^2_b (\mathbf{W}_{BB,b})^H (\mathbf{W}_{RF,b})^H \mathbf{W}_{RF,b} \mathbf{W}_{BB,b},\\
\mathbf{R}_{n,e} &= &\sigma^2_e (\mathbf{W}_{BB,e})^H (\mathbf{W}_{RF,e})^H \mathbf{W}_{RF,e} \mathbf{W}_{BB,e}.  \vspace{-0.2 cm}
\end{eqnarray}

\begin{figure}[t]
\centering
\includegraphics[width= 2.8 in]{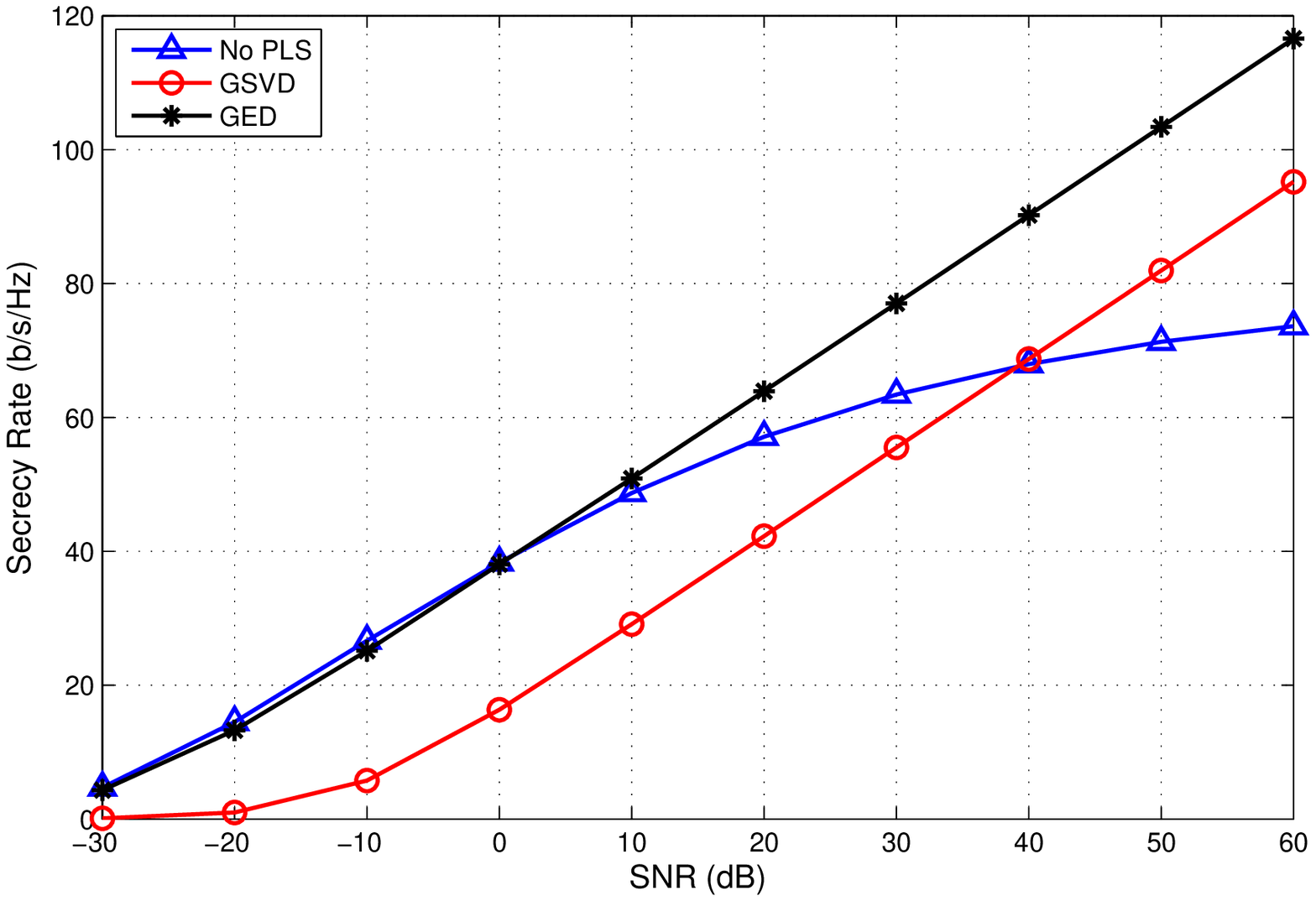} \vspace{-0.3 cm}
\caption{Secrecy rate versus SNR, \textit{full-digital} precoder and combiner ($N_a=N_b=N_e=192$, $N_{RF}=4$).}
\label{fig:full digital}\vspace{-0.0 cm}
\includegraphics[width= 2.8 in]{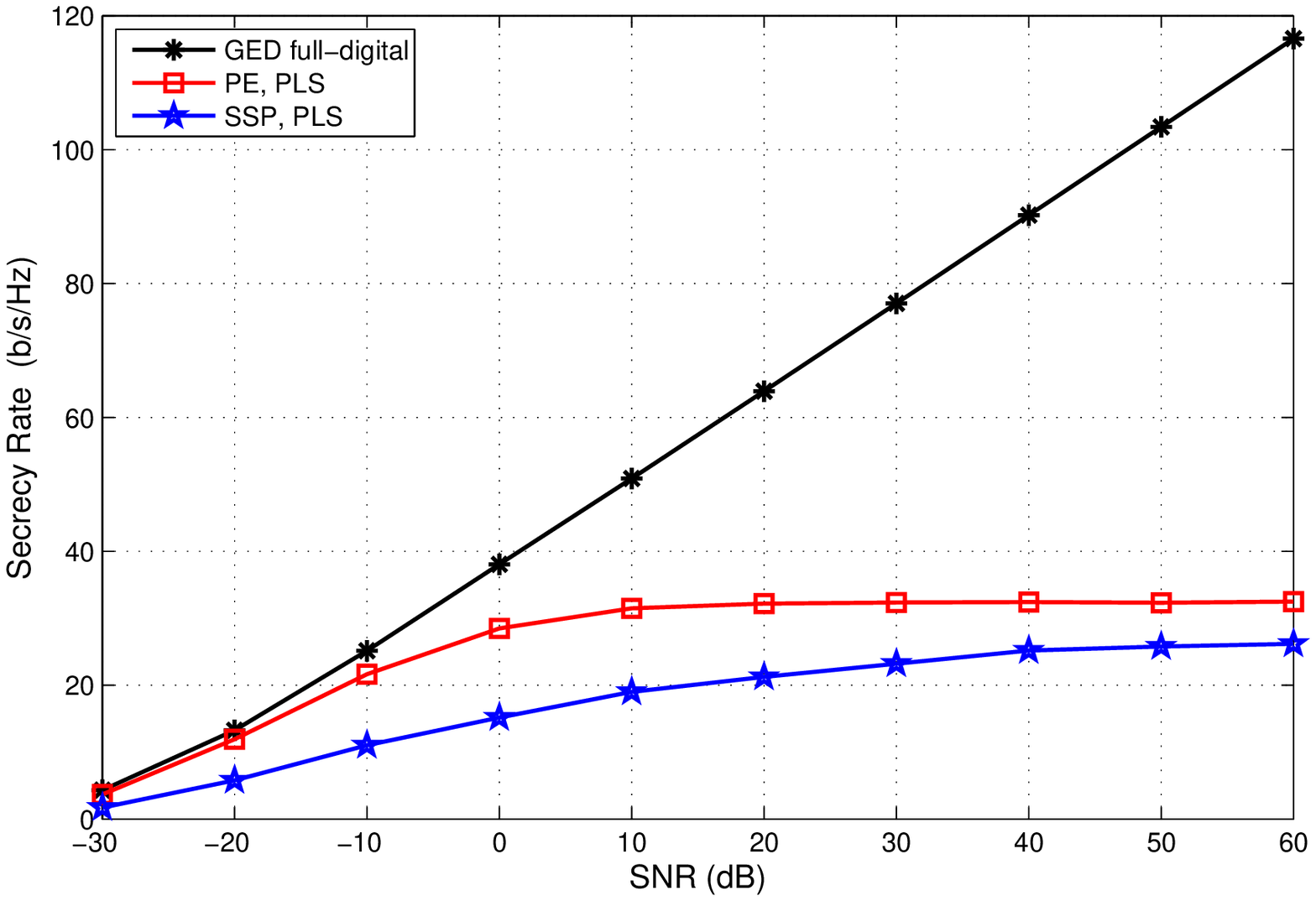}\vspace{-0.3 cm}
\caption{Secrecy rate versus SNR, \textit{hybrid} precoder and combiner  ($N_a=N_b=N_e=192$, $N_{RF}=4$).}
\label{fig:hybrid gap}\vspace{-0.3 cm}
\end{figure}

In the following, we carry out simulations to illustrate the security threaten in the mmWave MIMO systems. We assume both $\mathbf{H}_b$ and $\mathbf{H}_e$ are known to Alice and Bob
and first consider the full-digital precoder and combiner scheme.  Fig. \ref{fig:full digital} shows the secrecy rate under beamforming designs with: 1) no PLS effort; 2) generalized singular value decomposition (GSVD)-based PLS approach \cite{GSVD}; 3) generalized eigen decomposition (GED)-based PLS approach \cite{GED}.
It can be verified that the mmWave MIMO systems with ordinary (no PLS effort) beamforming design have notable information leakage in the high SNR range. Therefore, the legitimate transceiver needs to adopt beamforming with PLS efforts.
In addition, unlike the conventional MIMO system,  the GSVD approach is not as good as the GED approach under mmWave MIMO systems, which is because of the sparsity of the mmWave channel.
Therefore, we will use GED effort with full-digital beamforming as our secrecy performance benchmark in the following simulation studies.

In order to spotlight the impact of hybrid precoding and combining architecture considered in this paper, in Fig. \ref{fig:hybrid gap} we conduct the simulation using two representative hybrid beamforming algorithms: 1) codebook-based Spatially Sparse Precoding (SSP) \cite{SSP}; 2) codebook-free PE-AltMin (PE) \cite{PE AltMin}.  
For the PLS effort, we use these two approaches to find secure hybrid beamformers by minimizing the Euclidean distance between hybrid beamformers
and the GED-based full-digital secure beamformers.  
Fig. \ref{fig:hybrid gap} illustrates that the secrecy rate decreases dramatically when mmWave MIMO systems employ hybrid precoder and combiner, resulting in severe information leakage to the eavesdropper.
Moreover, the gap between the hybrid beamforming algorithms and the full-digital benchmark is quite large, which awaits researchers' investigation.

Inspired by the phenomenon illustrated in Figs. \ref{fig:full digital} and \ref{fig:hybrid gap}, in this paper we aim to develop a PLS-based hybrid precoder and combiner design for the secure transmission. Specifically, let $\mathcal{F}$ and $\mathcal{W}$ denote the beamsteering codebooks for the analog precoder and combiner, respectively. If $B^{RF}_t$ ($B^{RF}_r$) bits are used to quantize the AoD (AoA), $\mathcal{F}$ and $\mathcal{W}$ will consist of all possible analog precoding and combining vectors, which can be presented as \vspace{-0.1 cm}
\begin{eqnarray}
\mathcal{F} & = &\{ \mathbf{a}_t(2 \pi i /2^{B_t^{RF}}) :   i=1, \ldots,   2^{B_t^{RF}}\}, \vspace{-0.1 cm}\\
  \mathcal{W} & = &\{ \mathbf{a}_r(2 \pi i /2^{B_r^{RF}}) :  i=1, \ldots,   2^{B_r^{RF}}\}.\vspace{-0.1 cm}
\end{eqnarray}

\nid The PLS-based hybrid precoder and combiner design problem can be formulated as follows: \vspace{-0.1 cm}
\begin{eqnarray}
\hspace{-0.3 cm} \{ \mathbf{F}^*_{RF}, \mathbf{F}^*_{BB},  \mathbf{W}_{RF,b}^*, \mathbf{W}_{BB,b}^* \} \hspace{-0.3 cm} &=& \hspace{-0.3 cm} \arg \max R_s  \label{eq: temp1} \\
\hspace{-0.3 cm}  \mathrm{s.t.}  ~~~~ \mathbf{F}_{RF}(:,l) \hspace{-0.3 cm}  &\in& \hspace{-0.3 cm} \mathcal{F}, \forall l =1, \ldots, N_{RF} ,  \\
\hspace{-0.3 cm}  \mathbf{W}_{RF,b}(:,l) \hspace{-0.3 cm}  &\in& \hspace{-0.3 cm} \mathcal{W}, \forall l =1, \ldots, N_{RF} , \\
\hspace{-0.3 cm} \|\mathbf{F}_{RF} \mathbf{F}_{BB}\|^2_F \hspace{-0.3 cm} &=& \hspace{-0.3 cm} N_s.   \label{eq: digital power}
\end{eqnarray}

In the next section, we divide the problem according to the availability of the eavesdropper's CSI.
We first consider the scenario under which Alice and Bob know
eavesdropper's channel $\mathbf{H}_e$ and develop our codebook-based hybrid beamformer design algorithm.
Then, we study the case where the
eavesdropper's CSI is not available and AN-aided methods are adopted in the hybrid beamforming design problem.

\section{Secure Joint Hybrid Beamformer and Combiner Design}
\label{sec: ourscheme}

\subsection{Known Eavesdropper's Channel}
\label{subsec: both known}
Under this condition, our algorithm starts with performing singular value decomposition (SVD) of $\mathbf{H}_e$ as
\begin{equation}
\mathbf{H}_e = \mathbf{U}_e \mathbf{\Sigma}_e \mathbf{V}_e^H 
\end{equation}
\nid where $ \mathbf{U}_e$ and $\mathbf{V}_e$ are unitary matrices, $\mathbf{\Sigma}_e$ is an $N_e \times N_a$  diagonal matrix of singular values arranged in a decreasing order. Due to the sparsity of the mmWave MIMO channel, $ \mathbf{H}_e $ can be represented as
\begin{equation}
\mathbf{H}_e = \mathbf{\widetilde{U}}_e \mathbf{\widetilde{\Sigma}}_e \mathbf{\widetilde{V}}_e^H 
\end{equation}
where $\mathbf{\widetilde{\Sigma}}_e$ is a diagonal matrix whose elements are the first $L_b$ nonzero singular values, $\mathbf{\widetilde{U}}_e$ and $\mathbf{\widetilde{V}}_e$ contain the most $L_b$ left columns of  $\mathbf{U}_e$ and $\mathbf{V}_e$, respectively.

In an effort to prevent eavesdropping from Eve, Alice should elaborately design her precoder to avoid the AoD components of $\mathbf{H}_e$ in order to minimize Eve's reception.
Thus, to implement the secure transmission, we propose to remove the AoD  components of $\mathbf{H}_e$ from $\mathbf{H}_b$ by \vspace{-0.1 cm}
\begin{equation}
\mathbf{H}_1 \triangleq \mathbf{H}_b (\mathbf{I} - \widetilde{\mathbf{V}}_e \widetilde{\mathbf{V}}_e^H).
\label{eq: initial H} \vspace{-0.1 cm}
\end{equation}
\nid By this operation, $\mathbf{H}_1$ contains only the AoD components  of $\mathbf{H}_b$ but almost no AoD component of $\mathbf{H}_e$.
After this initial processing, we successively  select the $i$-th ($ i=1,\ldots, N_{RF}$) analog precoder and combiner pair to maximize
the corresponding channel gain while suppressing the co-channel
interference. The joint design problem can be successively solved by the following optimization problem: \vspace{-0.1 cm}
\begin{equation}
\{ \mathbf{w}^*_i, \mathbf{f}^*_i \} =\arg \underset{\substack{\mathbf{w}_{i}  \in \mathcal{W} \\ \mathbf{f}_{i}  \in \mathcal{F}}} {\textrm{max}} | \mathbf{w}^H_i \mathbf{H}_i \mathbf{f}_i | , i=1,\ldots, N_{RF}, \vspace{-0.1 cm}
\end{equation}
\nid and then assign them to the analog procoder and combiner matrices \vspace{-0.1 cm}
\begin{eqnarray}
\mathbf{F}^*_{RF}(:,i) &=& \mathbf{f}^*_i,\\
\mathbf{W}_{RF,b}^*(:,i) &=& \mathbf{w}^*_i. \vspace{-0.1 cm}
\end{eqnarray}

\nid Particularly, before executing the next iteration, we need to remove
the components of previous determined precoders and combiners
from the other data streams' channels
such that similar analog precoders and combiners will not be
selected by two different data streams. To achieve this goal, we let
$\mathbf{p}_i$ and $\mathbf{q}_i$  be the components of the
determined analog precoder and combiner for the $i$-th data
stream, respectively.
When $i=1$, $\mathbf{p}_1=\mathbf{f}^*_1$ and $\mathbf{q}_1=\mathbf{w}^*_1$; when $i>1$, the orthogonormal component $\mathbf{p}_i$ and $\mathbf{q}_i$ can be obtained by a Gram-Schmidt procedure: \vspace{-0.1 cm}
\begin{eqnarray}
 \mathbf{p}_i & = & \mathbf{f}_{i}^* - \sum\limits_{j=1}^{i-1}\mathbf{p}^H_j  \mathbf{f}_{i}^* \mathbf{p}_j, \nonumber \\ \mathbf{p}_i & = & \mathbf{p}_i/\|\mathbf{p}_i\|, i=2,\ldots,N_{RF};\\
\mathbf{q}_i & = & \mathbf{w}_{i}^* - \sum\limits_{j=1}^{i-1}\mathbf{q}^H_j\mathbf{w}_{i}^{\star}\mathbf{q}_j , \nonumber \\   \mathbf{q}_i & = &\mathbf{q}_i/\|\mathbf{q}_i\|, i=2,\ldots,N_{RF}.
\end{eqnarray}
Then $\mathbf{H}_{i+1}$ is updated for the next iteration by:
\begin{equation}
\mathbf{H}_{i+1} = (\mathbf{I}_{N_b} - \mathbf{q}_i \mathbf{q}^H_i) \mathbf{H}_i (\mathbf{I}_{N_a} - \mathbf{p}_i \mathbf{p}^H_i). \vspace{-0.1 cm}
\end{equation}

After determining the analog precoder $\mathbf{F}_{RF}^*$ and combiner $\mathbf{W}_{RF,b}^*$, we can obtain the effective channel $\mathbf{H}_{\mathrm{eff}} \triangleq (\mathbf{W}_{RF,b}^*)^H \mathbf{H}_b \mathbf{F}_{RF}^*$.
Then, an SVD-based baseband digital precoder is employed to further suppress the interference and maximize the sum-rate:
\begin{eqnarray}
 \mathbf{F}^*_{{BB}}  =  \mathbf{\bar{V}}(:,1:N_s), \label{eq:FBB} \\
 \mathbf{W}^*_{{BB,b}}  =  \mathbf{\bar{U}}(:,1:N_s), \label{eq:WBB}
\end{eqnarray}
where $\mathbf{H}_{\mathrm{eff}} = \mathbf{\bar{U}} \mathbf{\bar{\Sigma}} \mathbf{\bar{V}}^H$.
Finally, we normalize the baseband precoder $\mathbf{F}^*_{{BB}}$ by
\begin{equation}
\mathbf{F}^\star_{{BB}}=\frac{\sqrt{N_s}\mathbf{F}^*_{{BB}}}{\|\mathbf{F}^*_{RF}\mathbf{F}^*_{{BB}}\|_F}.
\label{eq:digital_normalize case1}
\end{equation}

\nid This secure hybrid precoder and combiner design algorithm is summarized in Table \ref{tb:a1}.

\begin{center}
\begin{table}[!t]  \vspace{0.0 cm}
\caption{Secure Hybrid Precoder and Combiner Design Algorithm with CSI of Eve.}\vspace{-0.2 cm}
\begin{center} \begin{small}
\begin{tabular}{l}
\hline \hline
\vspace{0.2 cm}\hspace{-0.2 cm} \textbf{Input:} $\mathcal{F}$, $\mathcal{W}$, $\mathbf{H}_1$.\\
\hspace{-0.2 cm} \textbf{Output:} $\mathbf{F}^*_{RF}$, $\mathbf{F}^*_{BB}$, $\mathbf{W}^*_{RF,b}$, and $\mathbf{W}^*_{BB,b}$.\\
\hspace{-0.2 cm} \textbf{for} $i=1:N_{RF}$ \\
\hspace{0.2 cm} $\left\{\mathbf{w}^*_i, \mathbf{f}^*_i  \right\}= \textrm{arg}\underset{\substack{\mathbf{w}_i \in \mathcal{W} \\ \mathbf{f}_i \in \mathcal{F}}} {\textrm{max}} | \mathbf{w}^H_i \mathbf{H}_i \mathbf{f}_i|;$\\
\hspace{0.2 cm} $\mathbf{F}^*_{RF}(:,i) = \mathbf{f}^*_i $;\\
\hspace{0.2 cm} $\mathbf{W}_{RF,b}^*(:,i) = \mathbf{w}^*_i $;\\
\hspace{0.2 cm} \textbf{if} $i=1$ \\
\hspace{0.5 cm} $\mathbf{p}_i=\mathbf{f}^*_i$, $\mathbf{q}_i=\mathbf{w}^*_i$. \\
\hspace{0.2 cm} \textbf{else}  \\
\hspace{0.5 cm} $\mathbf{p}_i= \mathbf{f}^\star_i -\sum\limits_{j=1}^{i-1}\mathbf{p}^H_i  \mathbf{f}^*_i \mathbf{p}_i$, $\mathbf{p}_i=\mathbf{p}_i/\|\mathbf{p}_i\|$;\\
\hspace{0.5 cm} $\mathbf{q}_i=\mathbf{w}^*_i -\sum\limits_{j=1}^{i-1}\mathbf{q}^H_j \mathbf{w}^*_i \mathbf{q}_j$, $\mathbf{q}_i=\mathbf{q}_i/\|\mathbf{q}_i\|$.\\
\hspace{0.2 cm} \textbf{end if}\\
\hspace{0.2 cm} $\mathbf{H}_{i+1} = (\mathbf{I}_{N_b} - \mathbf{q}_i \mathbf{q}^H_i) \mathbf{H}_i (\mathbf{I}_{N_a} - \mathbf{p}_i \mathbf{p}^H_i)$.\\
\hspace{-0.2 cm} \textbf{end for}\\
Obtain $\mathbf{F}^*_{BB}$ and  $\mathbf{W}^*_{BB,b}$
  by (\ref{eq:FBB})-(\ref{eq:digital_normalize case1}). \\
\hline \hline
\vspace{-0.0 cm}
\end{tabular}\label{tb:a1}\vspace{-0.5cm}
\end{small}
\end{center}
\end{table}
\end{center}

\subsection{Unknown Eavesdropper's Channel}

Under this condition, by common intuition, low-power Alice-to-Bob transmission can improve the security by making the signal interception of Eve more difficult.
Assume the Alice-to-Bob transmission needs to satisfy the QoS threshold $R_\gamma$, i.e. $R_b \geq R_\gamma$, $R_b=\log_2\det(\mathbf{I}_{N_s}+\mathbf{S}_b)$.
Thus, we can utilize the proposed algorithm in Table I with initialization $\mathbf{H}_1 = \mathbf{H}_b$ to find the optimal precoder $\mathbf{F}^*_{RF}$, $\mathbf{F}^*_{BB}$ and combiner $\mathbf{W}^*_{RF,b}$, $\mathbf{W}^*_{BB,b}$.
Then, we can find the minimum transmit power $P_s$ such that $R_b \geq R_\gamma$ can be satisfied.
To further increase the security, the residual power $P_{AN} \triangleq (P - P_s)^+$ is utilized for generating AN.

In this AN-based secure transmission, we assume $N_{RF}>N_s$ and the transmit signal becomes
\begin{equation}
\mathbf{x} = \mathbf{F}_{RF}(\sqrt{P_s}\mathbf{F}_{BB} \mathbf{s} + \sqrt{P_{AN}}\mathbf{F}_{BB,w} \mathbf{w} ),
\end{equation}
where $\mathbf{w}$ represents the $(N_{RF}-N_s) \times 1$ artificial noise vector, $\mathbb{E}\{ \mathbf{w}\mathbf{w}^H \} = \frac{1}{N_{RF}-N_s }\mathbf{I}_{N_{RF}-N_s}$, $\mathbf{F}_{BB,w} $ is the digital precoder associated with the AN $\mathbf{w}$.
The idea behind the AN-based PLS approach is that the generated AN should not degrade the reception of Bob. To ensure this principle with the obtained optimal combiners $\mathbf{W}_{BB,b}^{*}$ and $\mathbf{W}_{RF,b}^{*}$, we should have \vspace{-0.1 cm}
\begin{eqnarray}
\mathbf{W}_{BB,b}^{*H} \mathbf{W}_{RF,b}^{*H} \mathbf{H}_b \mathbf{F}_{RF}^* \mathbf{F}_{BB,w} & =& \mathbf{0} \\
\Rightarrow \mathbf{W}_{BB,b}^{*H} \mathbf{H}_{\mathrm{eff}}  \mathbf{F}_{BB,w} & = & \mathbf{0} \vspace{-0.1 cm}
\end{eqnarray}
\nid where $\mathbf{H}_{\mathrm{eff}} \triangleq \mathbf{W}_{RF,b}^{*H} \mathbf{H}_b \mathbf{F}_{RF}^*$ and $\mathbf{H}_{\mathrm{eff}} = \mathbf{\bar{U}}  \mathbf{\bar{\Sigma}} \mathbf{\bar{V}}^H$ by SVD.
If $N_s < N_{RF}$, with the optimal  $\mathbf{W}_{BB,b}^*$ obtained by $ \mathbf{W}^*_{{BB,b}}  =  \mathbf{U}(:,1:N_s)$ as in (\ref{eq:WBB}),
  the digital precoder $\mathbf{F}_{BB,w}$ for the AN should have a form of \vspace{-0.1 cm}
\begin{equation}
\mathbf{F}_{BB,w}^* = \mathbf{V}(:, N_s+1: N_{RF}). \vspace{-0.1 cm}
\end{equation}
Finally, we normalize the AN digital precoder as \vspace{-0.1 cm}
\begin{equation}
\mathbf{F}^\star_{BB,w}=\frac{\mathbf{F}^*_{BB,w}}{\|\mathbf{F}^*_{RF}\mathbf{F}^*_{BB,w}\|_F}.
\label{eq:digital_normalize} \vspace{-0.1 cm}
\end{equation}

\nid This AN-based secure hybrid precoder and combiner design algorithm is summarized in Table \ref{tb:a3}.

\section{Simulation Studies}
\label{sec: sim}

In this section, we illustrate the simulation results of the proposed  hybrid precoder and combiner design for secure transmission in mmWave MIMO systems.
Consider a mmWave MIMO wiretap system in which Alice, Bob and Eve are all equipped with $N_a=N_b=N_e=192$ ULA antennas. The antenna spacing of all ULAs is $d=\frac{\lambda}{2}$.
The AoA/AoD is assumed to be uniformly distributed in $[0,2\pi]$. We assume there exists 20 scatterers from which Bob and Eve randomly select $3\sim8$ scatterers as propagation paths with a certain probability that Bob and Eve may share some common scatterers.
For simplicity, the noise variance $\sigma^2_b$ and $\sigma^2_e$ are set to 1. 
The codebooks consisting of array response vectors as (\ref{eq: codebook f}) and (\ref{eq: codebook w}) with 128 angle resolutions are uniformly quantized in $[0,2\pi]$.

\begin{figure}[t]
\centering
\includegraphics[width= 3.5 in]{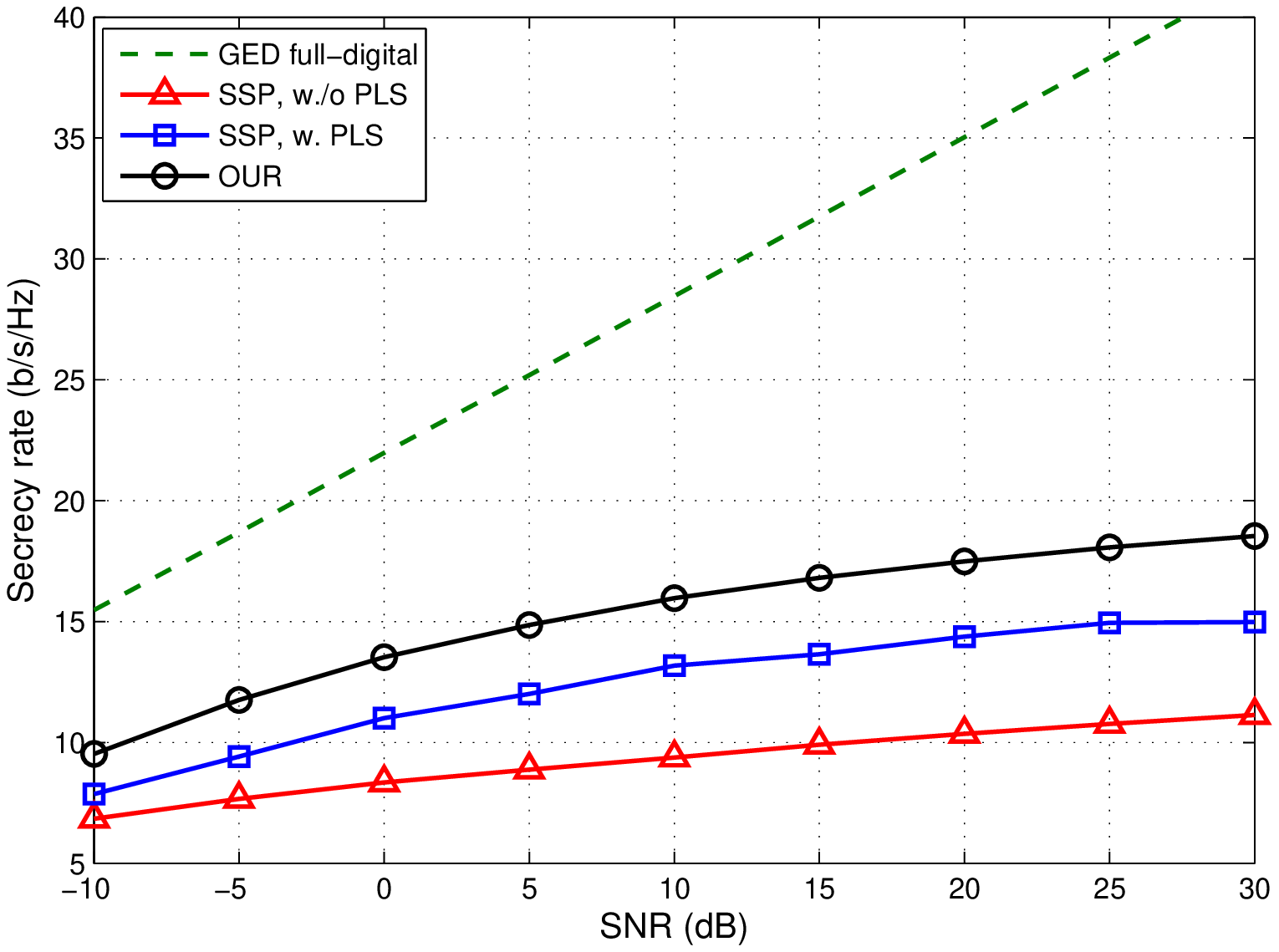}\vspace{-0.3 cm}
\caption{Secrecy rate versus SNR ($N_a=N_b=N_e=192$, $N_{RF}=2$, and $N_{s}=2$).}
\label{fig:case1 Cs 2}
\includegraphics[width= 3.5 in]{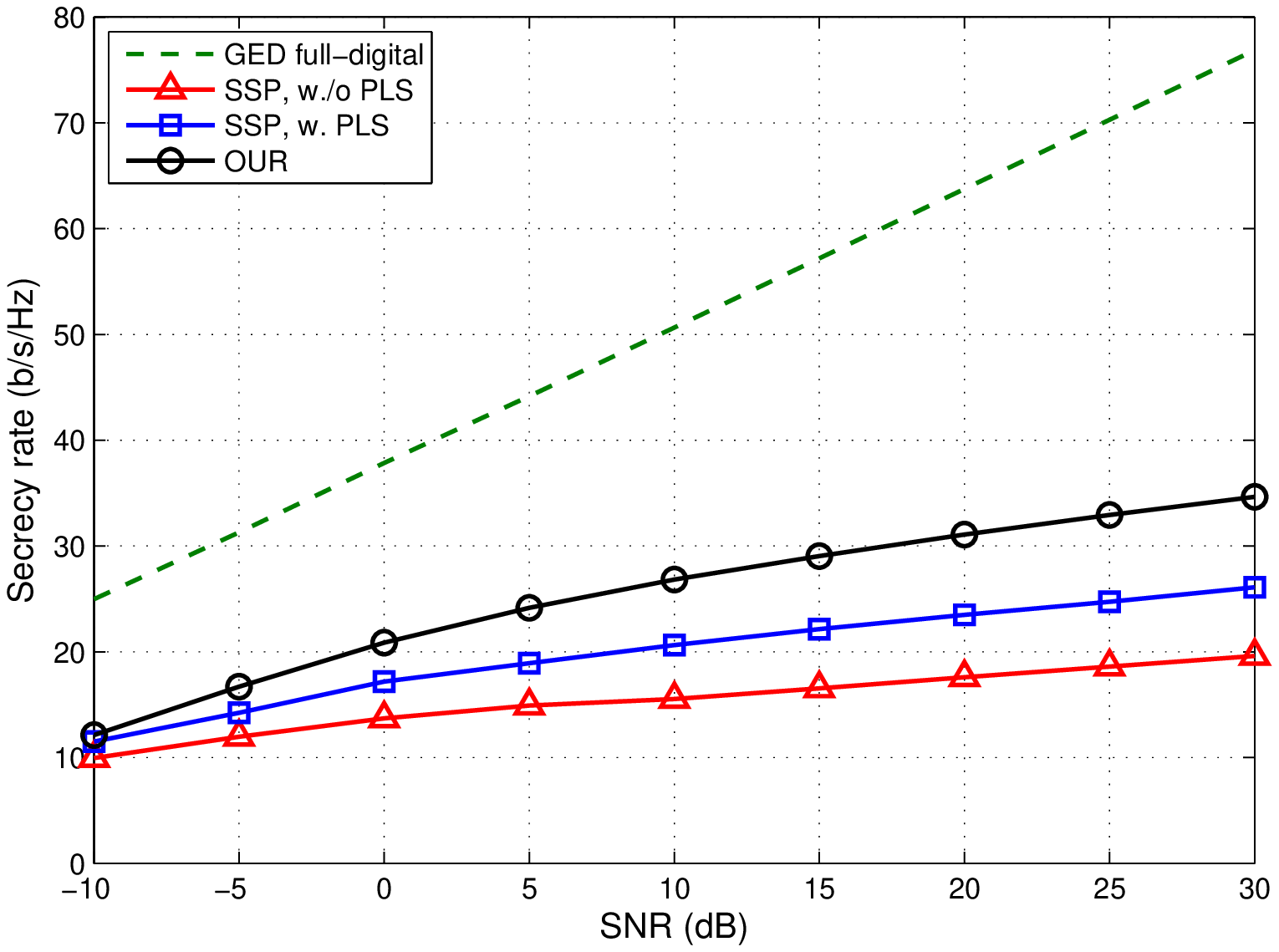}\vspace{-0.3 cm}
\caption{Secrecy rate versus SNR ($N_a=N_b=N_e=192$, $N_{RF}=4$, and $N_{s}=4$).}
\label{fig:case1 Cs 4}\vspace{-0.3 cm}
\end{figure}

We first consider the case that Eve's CSI is known and evaluate the proposed secure hybrid beamforming design algorithm in Table I.
For the comparison purpose, the representative codebook-based Spatially Sparse Precoding (SSP) algorithm with PLS effort and without PLS effort is also studies. For the fairness, we only focus the codebook-based algorithm and will not consider the codebook-free algorithms.
The full-digital beamforming design is also included as the performance benchmark. Fig. \ref{fig:case1 Cs 2} shows secrecy rate versus SNR for different beamforming design algorithms. The number of RF chains is set as $N_{RF}=2$, and the number of data streams is also $N_s=2$.
It can be observed from Fig. \ref{fig:case1 Cs 2} that our proposed algorithm can significantly outperform SSP algorithm in terms of secrecy rate. This result illustrates that our proposed algorithm can more efficiently prevent eavesdropping in the mmWave MIMO systems.
However, we also notice that our algorithm still has a notable gap between full-digital benchmark, which inspires us to pursue a more efficient hybrid algorithm in the future. In Fig. \ref{fig:case1 Cs 4}, a similar simulation is carried out with larger number of RF chains  $N_{RF}=4$ and larger number of data streams  $N_s=4$. The similar conclusion can be drawn.

\begin{center}
\begin{table}[!t]  \vspace{0.0 cm}
\caption{Secure Hybrid Precoder and Combiner Design Algorithm without CSI of Eve.}\vspace{-0.2 cm}
\begin{center} \begin{small}
\begin{tabular}{l}
\hline \hline
\hspace{-0.2 cm}  \textbf{Input:} ~ $\mathcal{F}$, $\mathcal{W}$, $\mathbf{H}_1 = \mathbf{H}_b$, $P$, $R_\gamma$.\\
\hspace{-0.2 cm} \textbf{Output:} $\mathbf{F}^*_{RF}$, $\mathbf{F}^*_{BB}$, $\mathbf{W}^*_{RF,b}$, $\mathbf{W}^*_{BB,b}$, $\mathbf{F}^*_{BB,w}$, $P_s$, and $P_{AN}$.\\
\hspace{-0.2 cm} \textbf{1.} ~Obtain optimal precoder $\mathbf{F}^*_{RF}$, $\mathbf{F}^*_{BB}$ and combiner \\
 \hspace{ 0.3 cm} $\mathbf{W}^*_{RF,b}$, $\mathbf{W}^*_{BB,b}$ using  Algorithm  in Table \ref{tb:a1}. \\
\hspace{-0.2 cm} \textbf{2.} ~Determine minimum $P_s$ such that $R_b \geq R_\gamma$. \\
\hspace{-0.2 cm} \textbf{3.} ~AN power $P_{AN} \triangleq (P - P_s)^+$.  \\
\hspace{-0.2 cm} \textbf{4.} ~AN digital precoder $\mathbf{F}_{BB,w}^* =  \mathbf{V}(:, N_s+1: N_{RF}) $.  \\
\hspace{-0.2 cm} \textbf{5.} ~Normalization  $\mathbf{F}^\star_{BB,w}=\frac{\mathbf{F}^*_{BB,w}}{\|\mathbf{F}^*_{RF}\mathbf{F}^*_{BB,w}\|_F}$.  \\
\hline \hline
\vspace{-0.0 cm}
\end{tabular}\label{tb:a3}\vspace{-0.5cm}
\end{small}
\end{center}
\end{table}
\end{center}

\begin{figure}[t]
\centering
\includegraphics[width= 3.5 in]{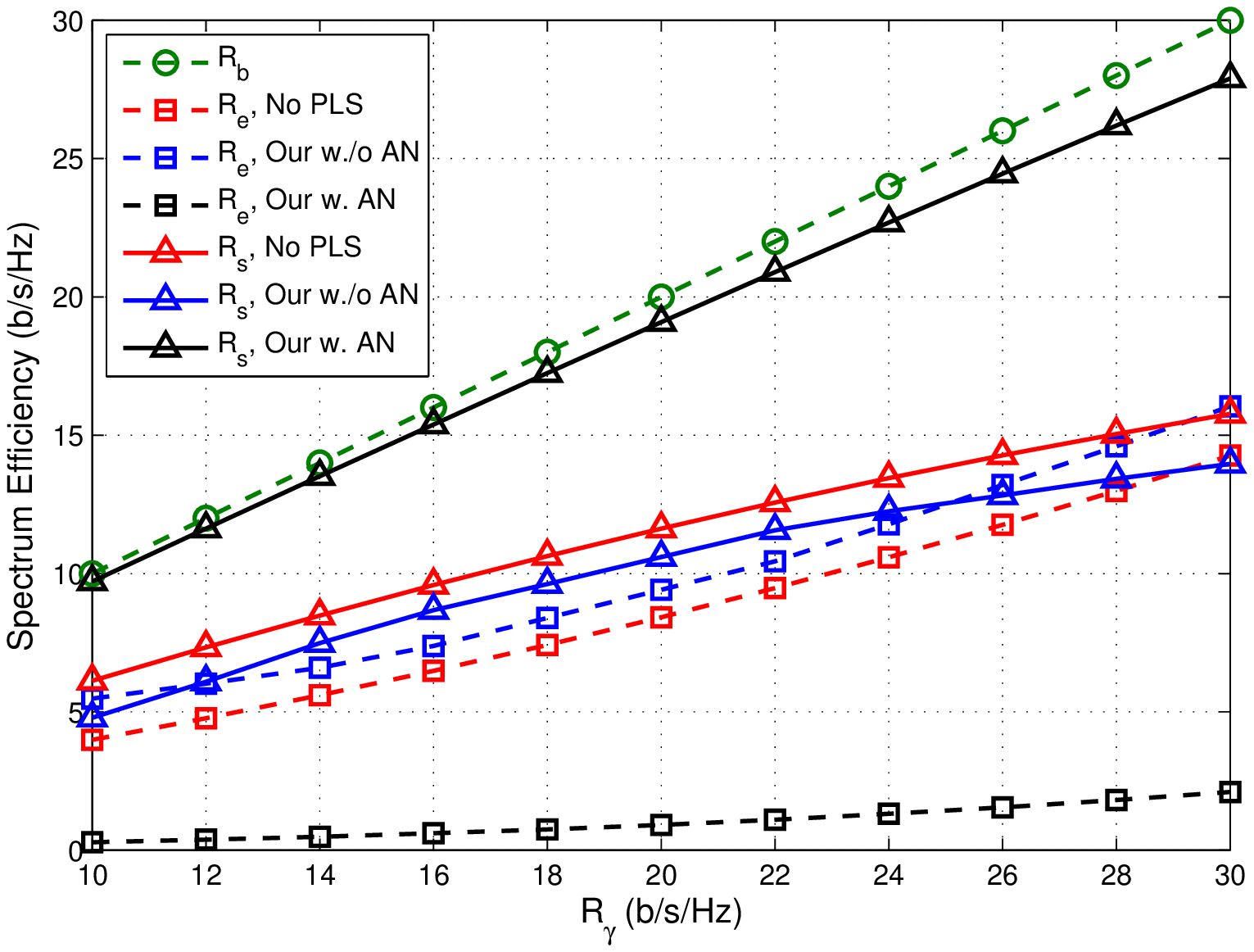}\vspace{-0.3 cm}
\caption{Spectrum efficiency versus $R_\gamma$ ($N_a=N_b=N_e=192$,   $N_{RF}=8$, and $N_s=4$).}
\label{fig:case2 Cs 8}\vspace{-0.0 cm}
\includegraphics[width= 3.5 in]{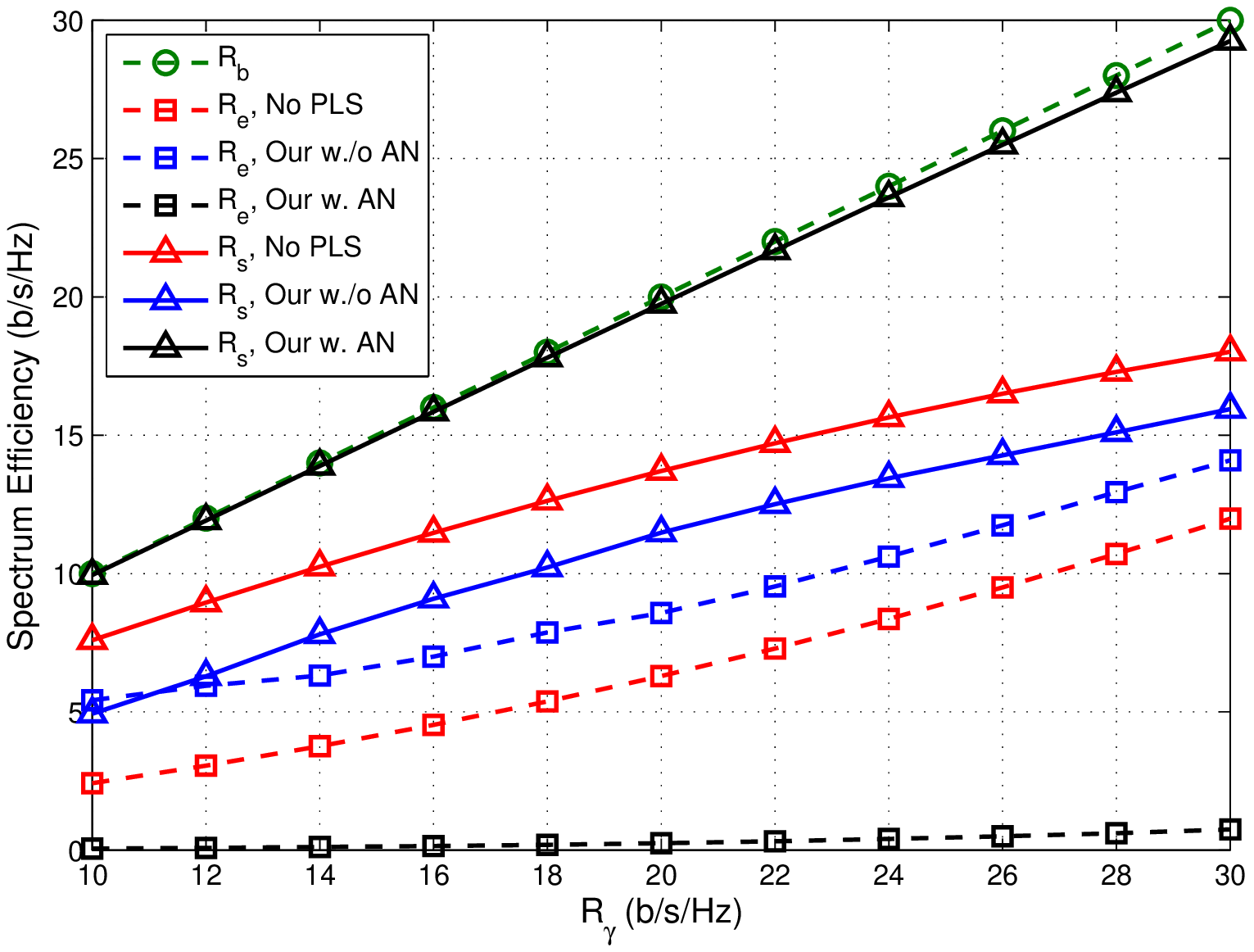}\vspace{-0.3 cm}
\caption{Spectrum efficiency versus $R_\gamma$ ($N_a=N_b=N_e=192$,  $N_{RF}=16$, and $N_s=4$).}
\label{fig:case2 Cs 16}\vspace{-0.3 cm}
\end{figure}

\vspace{-0.3 cm}
Next, we focus on the case that the eavesdropper's CSI is unknown. The number of RF chains is set as $N_{RF}=8$, and the number of data streams is $N_{s}=4$.
Fig. \ref{fig:case2 Cs 8} shows the spectrum efficiency of Bob $R_b$, spectrum efficiency of Eve $R_e$, and the resulting secrecy rate $R_s = (R_b-R_e)^+$ with the required legitimate transmission QoS $R_\gamma$.
It can be clearly observed that our proposed AN-based algorithm can dramatically reduce $R_e$ by adding AN in the transmit signaling while maintaining the required QoS $R_b$. Therefore,  our proposed AN-based hybrid beamforming design can implement secure transmission even when the CSI of the passive eavesdropper is unknown. In Fig. \ref{fig:case2 Cs 16}, we repeat the similar simulation by increasing the number of RF chains to $N_{RF}=16$. The secrecy performance becomes better since more spatial DoF can be utilized for generating AN to interrupt Eve's reception.

\section{Conclusion}
\label{sec: conclusion}

In this paper, we proposed a hybrid precoder and combiner design for secure transmission in a mmWave MIMO wiretap system.
With the eavesdropper's CSI, a joint analog precoder and combiner design algorithm was proposed to prevent the information leakage to the eavesdropper. When eavesdropper's CSI is unknown, we developed an AN-based hybrid beamforming approach, which can jam the eavesdropper's reception while maintaining the required QoS of the intended receiver.
Simulation results demonstrated the significant secrecy performance improvement of our propose algorithms compared with other hybrid beamforming algorithms.


\vspace{-0.0 cm}

\begin{thebibliography}{99}


\bibitem{Rappaport IA 13} T. S. Rappaport, S. Sun, R. Mayzus, H. Zhao, Y. Azar, K. Wang, G. N. Wong, J. K. Schulz, M. Samimi and F. Gutierrez ``Millimeter wave mobile communications for 5G cellular: It will work!'' \textit{IEEE Access}, vol. 1, pp. 335-349, May 2013.


\bibitem{Wang CL} Z. Wang, M. Li, X. Tian, and Q. Liu, ``Iterative hybrid precoder and combiner design for mmWave multiuser MIMO systems,'' \textit{IEEE Commun. Lett.}, accepted to appear, 2017.


\bibitem{SSP} O. E. Ayach, S. Rajagopal, S. Abu-Surra, Z. Pi, and R. W. Heath Jr., ``Spatially sparse precoding in millimeter wave MIMO systems,'' \textit{IEEE Trans. Wireless Commun.}, vol. 13, no. 3, pp. 1499-1513, Mar. 2014.

%


    \bibitem{Alkhateeb TCOM 16} A. Alkhateeb and R. W. Heath Jr., ``Frequency selective hybrid precoding for limited feedback millimeter wave systems,'' \textit{IEEE Trans. Commun.}, vol. 64, no. 5, pp. 1801-1818, May 2016.


\bibitem{Gao TVT 16} X. Gao, L. Dai, C. Yuen, and Z. Wang, ``Turbo-like beamforming based on Tabu search algorithm for millimeter-wave massive MIMO systems,'' \textit{IEEE Trans. Veh. Technol.}, vol. 65, no. 7, pp. 5731-5737, July 2016.


\bibitem{PE AltMin} X. Yu, J.-C. Shen, J. Zhang, and K. B. Letaief, ``Alternation minimization algorithms for hybrid precoding in millimeter wave MIMO systems,'' \textit{IEEE J. Sel. Topics Signal Process.}, vol. 10, no. 3, pp. 485-500, April 2016.


\bibitem{Gao JSAC 16} X. Gao, L. Dai, S. Han, C.-L. I, and R. W. Heath Jr., ``Energy-efficient hybrid analog and digital precoding for mmWave MIMO systems with large antenna arrays,'' \textit{IEEE J. Sel. Areas Commun.}, vol. 34, no. 4, pp. 998-1009, April 2016.


%

\bibitem{GED} A. Khisti, and G. W. Wornell, ``Secure transmission with multiple antennas I: the MISOME wiretap channel,'' \textit{IEEE Trans. Inf. Theory}, vol. 56, no. 7, pp. 3088-3104, July 2010.

\bibitem{GSVD} A. Khisti, and G. W. Wornell, ``Secure transmission with multiple antennas-part II: the MIMOME wiretap channel,'' \textit{IEEE Trans. Inf. Theory}, vol. 56, no. 11, pp. 5515-5532, Nov. 2010.


\bibitem{Wanghm AN2}  C. Wang and Hui-Ming Wang, ``Physical layer security in millimeter wave cellular networks,'' \textit{IEEE Trans. Wireless Commun.}, vol. 15. no. 8, pp. 5569-5585, Aug. 2016.


\bibitem{Wanghm BF} Y. Ju, H.-M. Wang, T.-X. Zheng, Q. Yang, Y. Zhang, Z. Li, P. Mu, and Q. Yin, ``Multi-antenna secure transmissions for millimeter wave wiretap channels,'' in \textit{Proc. IEEE Global Telecommunications Conference
Workshop on Trusted Communications with Physical Layer Security
(Globecom¡¯16 - Workshop - TCPLS)}, Washington, D.C., Dec. 2016.


 \bibitem{Wanghm AN1} Y. Ju, H.-M. Wang, and T.-X. Zheng, ``Secure transmissions in millimeter wave systems,'' \textit{IEEE Trans. Commun.}, accepted to appear, 2016.



\bibitem{scatter sharing channel} T. S. Rappapport, G. R. MacCartney, M. K. Samimi Jr., and S. Sun, ``Wideband millimeter-wave propagation measurements and channel models for future wireless communication system design,'' \textit{IEEE Trans. Commun.}, vol. 63, no. 9, pp. 3029-3056, Sept. 2015.



%


%


\end{thebibliography}
\end{document}